\documentclass[%
 reprint,
nofootinbib,
 amsmath,amssymb,
 aps]{revtex4-2}

\usepackage{graphicx}
\usepackage{dcolumn}
\usepackage{bm}

\usepackage{comment}
\usepackage{amsmath}

\DeclareUnicodeCharacter{2212}{-}

\newcommand{\cR}{\mathcal{R}}

\newcommand{\p}{\partial}

\newcommand{\be}{\begin{equation}}
\newcommand{\ee}{\end{equation}}

\newcommand{\benl}{\begin{equation*}}
\newcommand{\eenl}{\end{equation*}}

\begin{document}

\title{Emergence of inflaton potential from asymptotically safe gravity}
 
\author{Agustín Silva}
\email{agustin.silva@ru.nl}
\affiliation{High Energy Physics Department, Institute for Mathematics, Astrophysics, and Particle Physics, Radboud University, Nijmegen, The Netherlands.}

\date{\today}

\begin{abstract}
The Asymptotic Safety Hypothesis for gravity relies on the existence of an interacting fixed point of the Wilsonian renormalization group flow, which controls the microscopic dynamics, and provides a UV completion of the theory. Connecting such UV completion to observable physics has become an active area of research in the last decades. In this work we show such connection within the framework of scalar-tensor models. More specifically, we found that cosmological inflation naturally emerges from the integration of the RG flow equations, and that the predicted parameters of the emergent effective potentials provide a slow-roll model of inflation compatible with current observations. Furthermore, the RG evolution of the effective action starting at the UV fixed point, provides a prediction for the initial value of the inflaton field.
\end{abstract}

\maketitle

\section{Introduction}

Cosmological inflation has become a cornerstone in current theoretical cosmology since its initial proposal in \cite{Guth:1980zm}. It provides a mechanism for generating the entire observable universe from a single causal region of spacetime, solving the horizon problem \cite{Guth:1980zm,Linde:1981mu}. In addition, it provides an origin for the primordial density fluctuations that serve as seeds for structure formation on larger scales \cite{Lyth:1998xn}. See \cite{Bassett:2005xm,Baumann:2009ds,ParticleDataGroup:2016lqr,Nojiri:2010wj,Nojiri:2017ncd,Bartolo:2004if,Kofman:1994rk,Carr:2020gox} for reviews on the topic, and for other consequences of the inflationary scenario. 

There are several proposals for its realization, the most studied of them being single-field inflationary cosmologies \cite{Mukhanov:1990me}.
With sufficient fine tuning, one can construct an effective field theory (EFT) describing the dynamics of this scalar degree of freedom (inflaton) such that its predictions match observations \cite{Planck:2018jri}, but such models often require ad hoc shapes of the potential for the inflaton field, as well as a fine tuning of the initial conditions of the field itself \cite{Brandenberger:2016uzh}.

A tantalizing idea would be that such EFTs can be derived from first principles, and that the initial conditions of the inflaton come as a prediction of such derivation. Despite how ambitious this might sound, in this work we will provide a fundamental derivation of the inflaton potential and its initial conditions, using only non-perturbative Wilsonian renormalization group techniques \cite{Wilson:1971bg}.

More in particular, we will work under the Asymptotic Safety (AS) hypothesis for gravity, first proposed by Weinberg, and later implemented in $4$ dimensions in \cite{Reuter:1996cp}. In a nutshell, the AS hypothesis for gravity is the assumption the there exist an interacting renormalization group fixed point in the gravitational sector, typically called a Non-Gaussian fixed point (NGFP). For this UV completion to be predictive, this fixed point must come with the property that one only has to fine tune a finite amount of parameters such that the theory is attracted to it, or in a more technical language, its UV-critical surface must be finite dimensional. For reviews on the topic see \cite{Percacci:2007sz,percacci2017introduction,Reuter:2019byg}. This provides a UV completion of the theory, where only a finite amount of measurements are needed to fully determine the theory. Predictions from such UV completions of gravity are given by the endpoints of the RG trajectories that start at the NGFP \cite{Wetterich:1989xg,Codello:2015oqa,Saueressig:2024ojx,Shaposhnikov:2012zz,Reuter:2005kb,Knorr:2018kog}.

A primary tool for computing Wilsonian RG flows is the Wetterich equation 
\cite{wetterich1993exact},

\begin{equation}
    k \p_k \Gamma_k =\frac{1}{2}\mathbf{Tr}\left[ 
    \left( \Gamma_k^{(2)} + \cR_k \right)^{-1} k \p_k \cR_k \right]\, ,
    \label{FRGE}
\end{equation}
which encodes the dependence of so called ``effective average action'' $\Gamma_k$, on the coarse-graining scale $k$. In this equation, $\mathbf{Tr}$ represents a functional trace over all the degrees of freedom of the theory, $\cR_k$ is an IR regulator, and $\Gamma_k^{(2)}$ is the second order functional derivative of $\Gamma_k$ with respect to the field degrees of freedom. $\Gamma_k$ contains information of the path integral modes of energy higher than the momentum cut-off scale $k$. Since it is out of the scope of this work to dive into the details of the Wetterich equation, we refer the reader to the extensive bibliography about it, and references therein \cite{morris1994exact,Dupuis:2020fhh,berges2002non}.
For the purpose of this work, we are interested in complete (approximate) solutions of \eqref{FRGE}, $k \mapsto \Gamma_k$, which interpolate between a NGFP in the limit $k\rightarrow \infty$ (asymptotic safety) and the effective action $\Gamma \equiv \lim_{k \rightarrow 0} \Gamma_k$ associated with observables. In this limit where the regulator scale vanishes, the effective average action becomes the full effective action, containing information about all the energy modes of the theory.
The resulting $\Gamma$ that comes out if this integration is the prediction we are looking for, and is how we will connect the Asymptotic Safety hypothesis with cosmological inflation.

In particular, in this work we solve the functional RG flow equations for a class of scalar-tensor models, integrating them from the NGFP realizing the Asymptotic Safety hypothesis, up to the limit of vanishing regulator ($k=0$), where we make predictions for the scalar field interactions.

This paper is divided as follows. In sec. \ref{section:model} we describe the used scalar-tensor model. In sec. \ref{section:preliminaries} we describe the necessary preliminaries, such as coordinates choices and the parameterization of the interaction potentials, that we will use for finding numerical solutions of the RG flow equations. In sec. \ref{section:numericalsol} we show the results of a numerical integration of the flow equations. In sec. \ref{section:emergenceinflation} we show how the previous results naturally give rise to cosmological inflation, compatible with observations. We also discuss some of the implications of such results. In sec. \ref{section:conclusions} we present our conclusions and forthcomings. 

\section{The model}\label{section:model}

More specifically, here we will study the (euclidean) RG flow of scalar-tensor EFTs of the form 
\begin{equation}
  \Gamma_k=\int d^{4}\tilde{x}\sqrt{g}(-\tilde{F}(k,\tilde{\varphi})\tilde{R} +\tilde{V}(k,\tilde{\varphi}) 
  +\frac{1}{2}\partial_{\mu}\tilde{\varphi}\partial^{\mu}\tilde{\varphi})\, ,
  \label{eq:scalartensoransatz}
\end{equation}
where $\tilde{F}$ and $\tilde{V}$ are undetermined, even functions of the scalar field, $g$ is the metric determinant and $R$ the Ricci scalar. These type of models fall into the category of Horndeski theories \cite{Horndeski:1974wa}, and have been largely studied, using particular potentials $\tilde{F}$ and $\tilde{V}$, in both early and late time cosmology. See for example \cite{Copeland:2006wr,Bezrukov:2007ep,Kobayashi:2011nu,Koyama:2015vza}, and references therein.

Since the flow equations for the potentials $\tilde{F}$ and $\tilde{V}$ have been determined using \eqref{FRGE} in a number of works, we refer the reader to the bibliography on such computations for details of their derivation \cite{Percacci:2015wwa,Narain:2009fy,Henz:2013oxa,Percacci:2003jz}. Such computations in general involve a choice for an IR regulator $cR_k$ and a gauge fixing procedure, typically implemented by means of Faddeev–Popov ghosts. There are multiple methods for the resolution of the Wetterich equation, an example of those is an expansion in curvature invariants using heat-kernel methods. Typically, the obtained flow equations differ in some of their properties depending on the regulator choice, or the gauge fixing \cite{Ohta:2016jvw}. Predictions coming from these type of calculations should be taken with a pinch of salt, and every result should eventually be contrasted with other calculations using different methods, to verify its robustness regarding the previously mentioned choices. 

Having said that, and given that to the best of our knowledge this is the first time this particular calculations are performed, we will stick to a particular set of flow equations, and leave the previously mentioned cross-checks for a future work.

 In this paper, we will take equations \eqref{eq:flowVPercacci} and \eqref{eq:flowFPercacci}, derived in \cite{Percacci:2015wwa}, and use them as a starting point. Notice that in \eqref{eq:flowVPercacci} and \eqref{eq:flowFPercacci} the dimensionless variables 
\begin{equation}
      \varphi\doteq \frac{\tilde{\varphi}}{k} \,\,\,\,\,\,\,\,\, F \doteq \frac{\tilde{F}}{k^{2}} \,\,\,\,\,\,\,\,\, V \doteq \frac{\tilde{V}}{k^{4}} 
      \, ,
      \label{eq:originaldimlessvariables}
 \end{equation}
are being used for convenience. In these equations, the flow is computed at constant $\varphi$, for all values of $k$.

\begin{widetext}
\begin{align}
    k\,V^{(1,0)}(k,\varphi )= \varphi  V^{(0,1)}(k,\varphi ) -4 V(k,\varphi )+\frac{1}{16 \pi ^2} +\frac{3 F^{(0,1)}(k,\varphi )^2+F(k,\varphi )}{32 \pi ^2 \left(3 F^{(0,1)}(k,\varphi )^2+F(k,\varphi ) \left(V^{(0,2)}(k,\varphi )+1\right)\right)}
    \label{eq:flowVPercacci}
\end{align}
\begin{align}
    k&\,F^{(1,0)}(k,\varphi )=\varphi  F^{(0,1)}(k,\varphi
   )-2 F(k,\varphi )+\frac{37}{384 \pi ^2} \nonumber  \\ 
    &\quad +\frac{F(k,\varphi ) \left(\left(3 F^{(0,1)}(k,\varphi )^2+F(k,\varphi )\right) \left(-3 F^{(0,2)}(k,\varphi )+3 V^{(0,2)}(k,\varphi )+1\right)+2 F(k,\varphi ) V^{(0,2)}(k,\varphi )^2\right)}{96 \pi ^2 \left(3 F^{(0,1)}(k,\varphi )^2+F(k,\varphi ) \left(V^{(0,2)}(k,\varphi )+1\right)\right)^2}
    \label{eq:flowFPercacci}
\end{align}
\end{widetext}

In \cite{Percacci:2015wwa}, it was found that these flow equations present the following shift-symmetric ($\varphi \to \varphi +c $) NGFP 
\begin{equation}
       F_{*} = \frac{41}{768 \pi^{2}} \,\,\,\,\,\,\,\,\, V_{*}  = \frac{3}{128 \pi^{2}} \, ,
       \label{eq:NGFP}
 \end{equation}
compatible with the Asymptotic Safety hypothesis for gravity.

 In order to asses the predictivity of the UV completion \eqref{eq:NGFP} of \eqref{eq:scalartensoransatz}, one must determine the number of parameters of the theory that are a priori undetermined by the RG flow. These parameters are often called ``relevant directions''.  Relevant directions are the perturbations of the fixed point that are attracted to it when one lets them evolve following the RF flow. Perturbations that are repelled from the fixed point are called irrelevant directions. If at first order in perturbations there are directions that are neither repelled nor attracted to the fixed point, they are called marginal directions. These directions often require an analysis at higher order in perturbations around the fixed point. A determination of such directions is carried out by making a stability analysis of the fixed point.

 In the case of the model \eqref{eq:scalartensoransatz}, to determine its relevant directions, the authors of \cite{Percacci:2015wwa} perturbed the flow equations \eqref{eq:flowVPercacci} and \eqref{eq:flowFPercacci} at linear order in deviations from the NGFP using the ansatz $F \simeq F_{*} + \delta F \, (\frac{k0}{k})^{\theta}$ and $V \simeq V_{*} + \delta V \, (\frac{k0}{k})^{\theta}$, where $k0$ is a finite value for the coarse-graining scale, and $\theta$ is the so called critical exponent. Directions with $\theta > 0$, $\theta < 0$ or $\theta = 0$ are considered relevant, irrelevant and marginal, respectively, due to their behavior when increasing $k$ to infinity. The authors of \cite{Percacci:2015wwa} determined $2$ directions that were relevant,
 \begin{equation}
   (\delta F,\delta V)_1=(0,1), \, \theta_{1}=4  \,\,\,\,\,\,\,\, (\delta F,\delta V)_2=(1,0), \, \theta_{2}=2 \, ,
   \label{eq:reldirgravity}
\end{equation}
and $1$ direction that at linear order seemed to be marginal
\begin{equation}
    (\delta F,\delta V)_3=(-1+32\pi^2\varphi^2,0), \, \theta_{3}=0 \,.
    \label{eq:reldirbreaksimmetry}
\end{equation}
All other directions were found to be irrelevant.

With this information, one might naively conclude that the only relevant directions are $(\delta F,\delta V)_1$ and $(\delta F,\delta V)_2$, and thus NGFP \eqref{eq:NGFP} is not interesting for scalar field phenomenology. This is because if that were the case, there would be no relevant direction that creates shift symmetry breaking in the scalar field direction, meaning that the Ward identities would forbid the emergence of non-trivial interactions for the scalar field \cite{Percacci:2016arh}. 

Because of this, we decided to extend the analysis of $(\delta F,\delta V)_3$ to second order, to determine its relevance, where we found that $(\delta F,\delta V)_3$ is also a relevant direction. The calculations can be found in the appendix \ref{sec:appendixreldirbs}. This means that at smaller energies, RG trajectories emerging from the NGFP \eqref{eq:NGFP}, can potentially have symmetry breaking, and create non-trivial interactions in the scalar field sector. Due to this, the fixed point \eqref{eq:NGFP} can have interesting phenomenology in the IR, and we will solve \eqref{eq:flowVPercacci} and \eqref{eq:flowFPercacci} using as initial conditions the UV completion given by \eqref{eq:NGFP}. 

\section{Preliminaries for numerical methods}\label{section:preliminaries}

Since equations \eqref{eq:flowVPercacci} and \eqref{eq:flowFPercacci} are non-linear, second order, partial differential equations, we decided that numerical methods are the best approach to find a solution that uses the NGFP \eqref{eq:NGFP} as initial condition at $k \to \infty$.

Despite the simplicity of \eqref{eq:flowVPercacci} and \eqref{eq:flowFPercacci} using the variables \eqref{eq:originaldimlessvariables}, these are not the best variables to find a numerical solution. First of all, for computational reasons, one must work with compact domain intervals, and therefore one must compactify the RG interval $k\in [0,\infty)$ to some finite interval. Second, it is not convenient to work with variables that require a scaling with $k$ once approaching the $k \to 0$ limit. One example of this are the dimensionless field variables $\varphi$, where if one wants to remove the regulator $k$, and is interested in what happens at finite scalar field $\tilde{\varphi}$, one has to deal with $\infty \cdot 0$ limits ($\lim_{k \to 0}\varphi \, k=\tilde{\varphi}$). 

Because of this, we will search for solutions of $F$ and $V$ that are parameterized as
\begin{equation}
     F(k,\phi) = \frac{1+f(g_k,\phi^{2})}{16 \pi g_k^2} \,\,\,\,\, V(k,\phi) =\frac{2\lambda(g_k) +v(g_k,\phi^{2})}{(16 \pi g_k^2)^{2}} \, ,
     \label{eq:splittingFandV}
 \end{equation}
where $g_k \equiv \frac{\tilde{g_k}}{k}$ is a dimensionless parameter, with its corresponding dimensionful version $\tilde{g}$, 
 $\lambda_k$ is a dimensionless coupling
, and the field variable $\phi$ is defined as 
\begin{equation}
  \phi \equiv g\varphi=\tilde{g}\tilde{\varphi}\, .
    \label{eq:dimlessfieldvariable}
\end{equation}
Notice that $\phi$, $f$ and $v$ are dimensionless without the need of multiplication by $k$. In this parameterization we separated the scalar field dependent part into $f$ and $v$, and the purely gravitational part is parameterized by $g$ and $\lambda$. For this splitting, we assume that $f(g_k,0)=v(g_k,0)=0$, thus at vanishing field we recover the gravitational action. Furthermore, notice that we exchanged the $k$ dependence by a dependence on the coupling $g_k$. This is because of the following: at the fixed point \eqref{eq:NGFP}, in this parameterization, we have the following result
\begin{align}
    \lim_{k\to \infty} & g_k \equiv g_{*}=\sqrt{\frac{48 \pi}{41}}\simeq  1.92 \, , \nonumber \\ 
    \quad \lim_{k\to \infty} & \lambda(g_k)\equiv \lambda_{*}=\sqrt{\frac{995328 \pi^{3}}{68921}}\simeq 21.2 \, , \nonumber \\ 
    \quad \lim_{k\to \infty} & \{f(g_k,\phi^2), \, v(g_k,\phi^2)\}\equiv\{f_*, \, v_* \}=\{0, \, 0\} \, ,
     \label{eq:NGFPsplit1}
\end{align}
that indicates that the field dependent parts vanish in the UV. 

With the splitting given by \eqref{eq:splittingFandV}, the two relevant directions \eqref{eq:reldirgravity} are parameterized by $g$ and $\lambda$. This means that $g$ and $\lambda$ are undetermined by the RG flow, and should be determined by measurements. In particular, we have that $\lim_{k \to 0}\frac{g_k}{k}=\lim_{k \to 0}\tilde{g}_k = \sqrt{\tilde{G_0}}$, where $\tilde{G_0}$ is the measured value of Newton's constant, and implying that $\lim_{k \to 0}g_k=0$. This means that using as an RG parameter the coupling $g$, one can compactify the RG domain from $k \in [0,\infty)$ to $g \in [0,\sqrt{\frac{48 \pi}{41}}]$, thus solving the issue of compact domain for numerical solutions. On top of that, we have that $\lim_{k \to 0} \lambda(g_k)=16 \pi \Lambda_0\,G_0$, where $\Lambda_0$ is the measured Cosmological constant. Since $\lim_{k \to 0} \lambda(g_k)\simeq 10 ^{-120}$, we will approximate it by $0$ when making inflationary predictions.

Furthermore, the previous observations mean that the variable $\phi$ has a well defined interpretation in the $k \to 0$ limit, since it satisfies $\lim_{k \to 0}\phi = \sqrt{\tilde{G_0}}\tilde{\varphi}$, meaning that at $k=0$ the variable $\phi$ is just the scalar field in Planck mass units. 

 The main goal of this separation is that now we can encapsulate all irrelevant directions inside $f$ and $v$, and we would like to make predictions for all those. In particular, we want to determine the functional form of $\lim_{k\to \infty} \{f(g_k,\phi^2), \, v(g_k,\phi^2)\} = \{f(0,\phi^2), \, v(0,\phi^2)\}$. In order to turn the RG flow equations \eqref{eq:flowFPercacci} and \eqref{eq:flowVPercacci} in terms of $f$ and $v$, one has to make a series of transformations. Due to their length, those transformations, and the resulting flow equations can be found in the appendix \ref{sec:appendixequations}.   
 
\begin{figure*}
  \includegraphics[width=\textwidth]{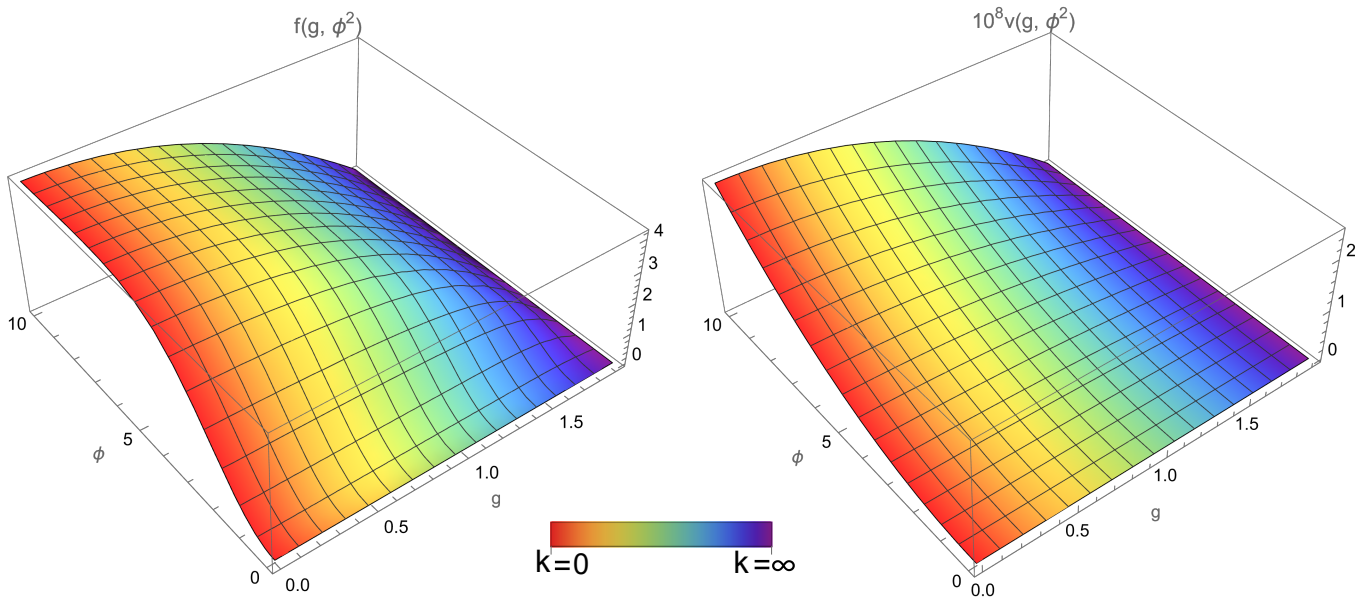}
  \caption{Numerical solutions for the functions $f(g,\phi^2)$ and $v(g,\phi^2)$ in the domain $(g,\phi)\in [0,\sqrt{\frac{48 \pi}{41}}]\times[0,10]$ using boundary conditions \eqref{eq:boundarycond1stderv} and \eqref{eq:boundarycond1stderf}, fixing $m_{0}^{f}= 0.42$ and $m_{0}^{v}= 2.75 \times 10^{-10}$. The predicted potentials can be extracted from the intersection of the surfaces with the plane $g=0$. The color coding is a linear function of $g$ that interpolates between red and purple, meant to represent small and high values of the original coarse-graining scale $k$. }
  \label{fig:scalingsolutions}
\end{figure*}

\section{Numerical solution of RG flow equations}\label{section:numericalsol}

The resulting flow equations for $f(g,\phi^2)$ and $v(g,\phi^2)$ are given by \eqref{eq:repflowv} and \eqref{eq:repflowf}, and they are coupled, non-linear, second-order differential equations. So far, the only constrains to be compatible with the NGFP \eqref{eq:NGFP} are that $f(g_*,\phi^{2})=v(g_*,\phi^{2})=0$, and due to the splitting used, we have $f(g,0)=v(g,0)=0$. As mentioned in section \ref{section:model}, there is one extra relevant direction given by \eqref{eq:reldirbreaksimmetry}. This direction will come with an associated coupling in the action that is undetermined by the RG flow. Besides, since this direction is field dependent, it is contained within $f$. This means we will have at least $1$ degree of freedom in the resulting $f$, after we integrate the RG flow equations. 

In order to find the number of free parameters that will be undetermined by the RG flow, and to see if there are any extra degrees of freedom inside $v$, it is not sufficient to look only at the stability properties around the fixed point (see section \ref{section:model}), but one must also look at differential order the equation determining the RG flow. In our case, since the flow equations are second order, but we only have the conditions $f(g_*,\phi^{2})=v(g_*,\phi^{2})=0$ and $f(g,0)=v(g,0)=0$, there is still one extra boundary condition to be fixed. This can add extra degrees of freedom to the solutions, apart from the relevant directions around the fixed point.

In order to close the system of equations, we will impose boundary conditions by using the derivatives $f^{(0,1)}(g,0)\equiv f_1(g)$ and $v^{(0,1)}(g,0)\equiv v_1(g)$. 

A priori one does not know which functions $f_1(g)$ and $v_1(g)$ are consistent with the equations and would give rise to a solution. In order to find candidates for these boundary conditions, we expand the flow equations \eqref{eq:repflowv} and \eqref{eq:repflowf} near $\phi=0$ up to order $\phi^2$, assuming $f(g,\phi^{2})=f_{1}(g)\phi^{2}$ and $v(g,\phi^{2})=v_{1}(g)\phi^{2}$. To determine $v_{1}(g)$ and $f_{1}(g)$ analytically, one can assume $|f_{1}(g)|,\, |v_{1}(g)|<1$, and expand \eqref{eq:repflowv} and \eqref{eq:repflowf} to the lowest non-trivial order in $f_1$ and $v_1$. The resulting equations are
\begin{align}
       82 g \,v_1(g)+ \left(48 \pi -41 g^2\right) v_1'(g)&=0 \, , \, \nonumber \\ 
    12  g \, f_1(g){}^2+ \left(96 \pi ^2 -82 \pi  g^2 \right) f_1'(g)&=0 \, ,
\end{align}
and the solutions are 
\begin{equation}
        v^{(0,1)}(g,0)=v_1(g)=\frac{41 \left(\frac{48 \pi }{41}-g^2\right) m_0^v}{48 \pi }\, ,
        \label{eq:boundarycond1stderv}
\end{equation}
and 
\begin{equation}
        f^{(0,1)}(g,0)=f_1(g)=\frac{m_0^f}{1-\frac{3}{41 \pi }\,m_0^f\,\log \left(\frac{48 \pi -41 g^2}{48 \pi }\right)} \, ,
        \label{eq:boundarycond1stderf}
\end{equation}
where $m_0^v,\, m_0^f$ are free, but smaller than $1$, parameters. These boundary conditions are compatible with the NGFP, since they satisfy $\lim_{g\to g_{*}}v_1(g)=\lim_{g\to g_{*}}f_1(g)=0$. Furthermore, one of the degrees of freedom given by this equation is $m_0^f$, which parameterizes the relevant direction \eqref{eq:reldirbreaksimmetry}, and the existence of the extra degree of freedom $m_0^v$ is a consequence of the differential order of the RG flow equations. These two free parameters $m_0^f$ and $m_0^v$ determine the ``infrared mass couplings'' of both potentials $f$ and $v$, since they satisfy $\lim_{g\to 0}f_1(g)=m_0^f $ and $\lim_{g\to 0}v_1(g)=m_0^v $. 

Notice that this method to determine the boundary conditions is not unique, and one can come up with other boundary conditions that are well motivated and are still compatible with the NGFP and the flow equations themselves. In this paper, we will restrict to these boundary conditions, but modifications of those can be found in the supplementary Mathematica notebook. 

Using these results, one can close the system of equations with the boundary conditions \eqref{eq:boundarycond1stderv} and \eqref{eq:boundarycond1stderf}, plus $f(g_*,\phi^{2})=v(g_*,\phi^{2})=f(g,0)=v(g,0)=0$. The numerical solution implementation can be found in the supplementary material, where we use Mathematica to find a solution for equations \eqref{eq:repflowf} and \eqref{eq:repflowv} on the range $\phi \in [0,10]$.

The numerical solutions for the particular choice $m_{0}^{f}= f^{(0,1)}(0,0)= 0.42$ and $m_{0}^{v}= v^{(0,1)}(0,0)= 2.75 \times 10^{-10}$, can be seen in Fig. \ref{fig:scalingsolutions}. These values are fine tuned in order to get the desired results in the next section, and different values would change the outcome. 
In a way, these two degrees of freedom are spanning the set of possible theories emerging from the NGFP. 

\begin{figure}
    \centering
    \includegraphics[width=\columnwidth]{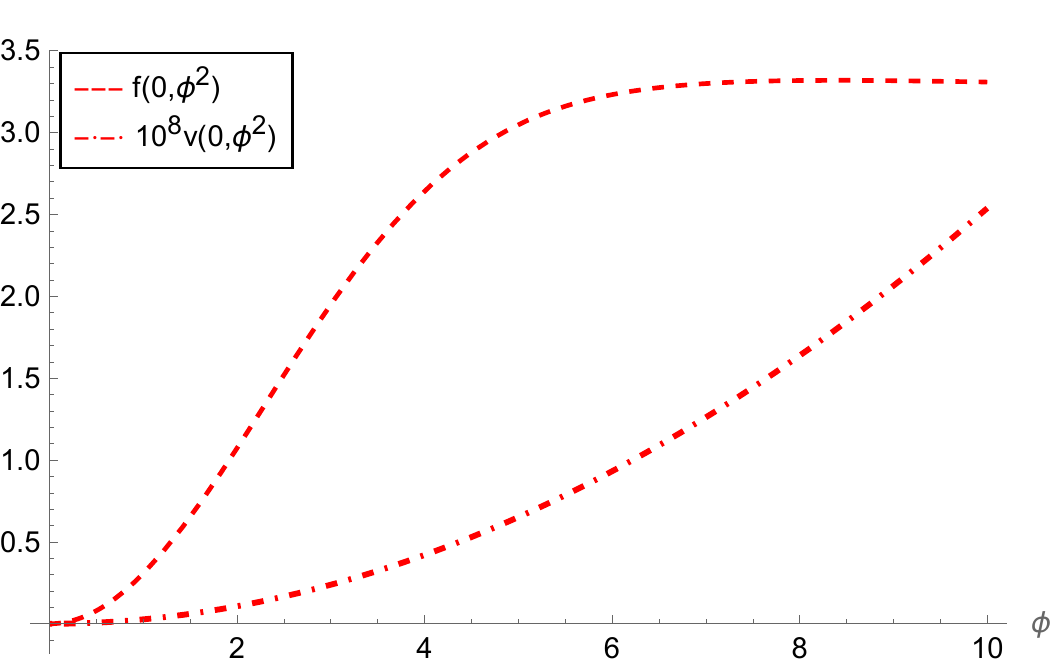}
    \caption{Numerical result for the predicted functional dependence of $f$ an $v$ at vanishing regulator ($k=0$), in the range $\phi \in [0,10]$, for the choices $m_{0}^{f}= 0.42$ and $m_{0}^{v}= 2.75 \times 10^{-10}$.}
    \label{fig:predictedfandv}
\end{figure}

In Fig. \ref{fig:scalingsolutions} we can see that the shift-symmetry of the NGFP is broken when $g<g_*$, i.e. when the coarse graining scale $k$ is less than infinity. This type of solutions are only allowed thanks to the symmetry breaking relevant direction \eqref{eq:reldirbreaksimmetry}, without which the potentials $f$ and $v$ will remain $0$ throughout the entire flow, preserving the shift-symmetry of the NGFP. 

The interaction potentials obtained at $k=0$, $f(0,\phi^2)$ and $v(0,\phi^2)$ can be extracted from Fig. \ref{fig:scalingsolutions}. Their functional dependence is depicted in Fig. \ref{fig:predictedfandv}. One can see that the predicted potentials $f(0,\phi^2)$ and $v(0,\phi^2)$ have non-trivial dependence with the scalar field in the studied range. These results are subject to a particular choice of the free parameters, and different choices give different results.
At least for the analysed values, one can see that the effect of changing the parameters $m_0^f$ and $m_0^v$ is an overall multiplication factor, and does not change the shape of the potentials (see appendix \ref{sec:appendixboundarycond}). 

Now, we are interested in the phenomenology of the predicted potentials. We found that this is easier to understand after a change of variables in the effective action, as we will show in the next section.

\section{Emergence of Inflaton Potential}\label{section:emergenceinflation}

The phenomenological consequences of the potentials obtained in Fig. \ref{fig:predictedfandv}, can be easily interpreted after a conformal change of variables. Specifically, the scalar-tensor model given in \eqref{eq:scalartensoransatz} is expressed in the so called ``Jordan frame'', but the same action can be transformed to the ``Einstein frame'' (E) \cite{Faraoni:1999hp,Postma:2014vaa}. To change to such frame, one performs a conformal transformation
\begin{equation}
    g^{E}_{\mu\,\nu}\equiv (1+f(g,\phi^{2}))g_{\mu\,\nu}
\end{equation}
such that the action \eqref{eq:scalartensoransatz} is turned into  
\begin{align}
  \Gamma_{k}=\int d^{4}x\sqrt{g^E}(-\frac{R^E}{16 \pi g^{2}} + \frac{V_{eff}(g,\phi(\sigma))}{(16 \pi g^{2})^{2}} 
  +\frac{\partial_{\mu}\sigma\partial^{\mu}\sigma}{2\,g^{2}}),
  \label{eq:ansatzactioneinsteinframe}
 \end{align}
where 
\begin{equation}
    V_{eff}(g,\phi(\sigma))\doteq\frac{2\lambda(g)+v(g,\phi^{2}(\sigma))}{(1+f(g,\phi^{2}(\sigma))^{2}},
    \label{eq:effPot}
\end{equation}
and where the new scalar field satisfies
\begin{equation}
    (\frac{\partial\, \sigma}{\partial\, \phi})^{2}=\frac{1}{(1+f(g,\phi^{2}))}+\frac{3\, (f^{'}(g,\phi^{2}))^{2}}{(1+f(g,\phi^{2}))^{2}}.
    \label{eq:relphisigma}
\end{equation}

\begin{figure}[ht]
    \centering
    \includegraphics[width=\columnwidth]{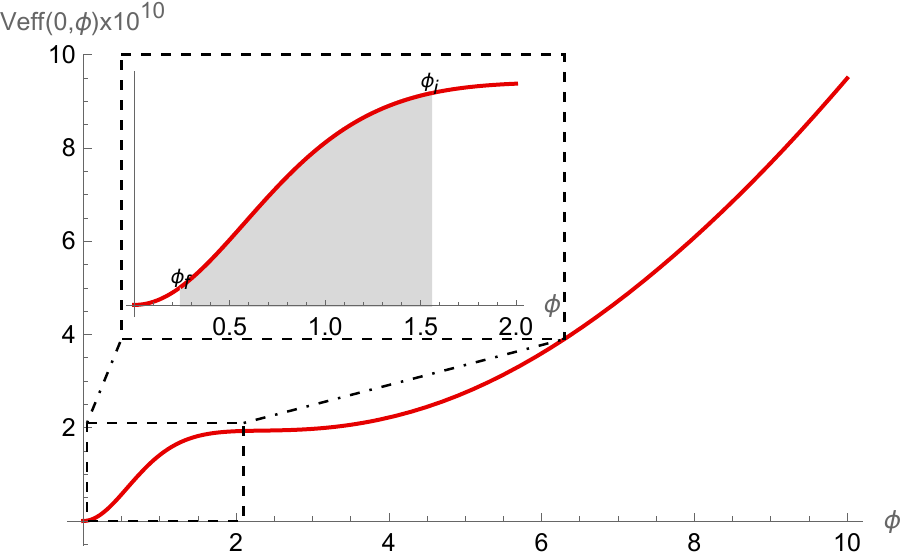}
    \caption{Obtained effective potential for the inflaton field as a function of $\phi$, in the range $\phi \in [0,10]$. The zoomed in region represents the region where a period of cosmological inflation, compatible with observations, can occur. The initial and final values of the field for generating such period are $\phi_i=1.55$ and $\phi_f=0.25$.}
    \label{fig:InflatonPotRegion}
\end{figure}

As we said before, we are interested in the phenomenology of the effective potentials once all quantum fluctuations are integrated out. This is obtained by looking at $V_{eff}(0,\phi(\sigma))$. In Fig. \ref{fig:InflatonPotRegion}
one can see the resulting effective potential as a function of the variable $\phi$. In this result we approximated $\lambda(0) \simeq 10^{-120} \simeq 0$, since we found it made no difference in the numerical predictions. Presenting the results as a function of $\phi$ is merely for simplicity, and the phenomenology is equivalent for the potential as a function of $\sigma$. If one wants to see the shape of the potential as a function of $\sigma$, one must solve \eqref{eq:relphisigma} numerically, at $g=0$. This can be found in the supplementary material.  

As one can immediately see in Fig. \ref{fig:InflatonPotRegion}, the potential presents an approximate plateau at values of $\phi$ between $1.5$ and $3$, which, as mentioned before, can be interpreted equivalently to values of the original scalar field $\tilde{\varphi}$ between $1.5$ and $3$ times the Planck mass.

This lead us to compute inflationary observables associated with a field starting at some value of such plateau. In the way the potential $V_{eff}$ is defined, the slow-roll observables \cite{Baumann:2022mni} take the form 
\begin{align}
    \epsilon & = \frac{1}{8 \pi}\frac{1}{2}(\frac{V_{eff}^{'}}{V_{eff}})^{2} \,\,\,\,\,\,\,\, \eta = \frac{1}{8 \pi}(\frac{V_{eff}^{''}}{V_{eff}}) \nonumber \\ 
    \quad n_s &= 1 - 6\,\epsilon + 2\,\eta \,\,\,\,\,\,\,\,\,\, r= 16\,\epsilon \nonumber \\ 
    \quad A_s &= \frac{1}{4} \frac{V_{eff}}{24 \pi^{2} \epsilon} \,\,\,\,\,\,\,\,\,\,\,\,\, N_{ef} = 8 \pi \int_{\sigma_{i}}^{\sigma_{f}}\frac{d\,\sigma}{\sqrt{2\,\epsilon}},
    \label{eq:slowrollparameters}
\end{align}
 where the prime denotes a derivative with respect to $\sigma$, $n_s$ is the spectral index, $r$ is the scalar to tensor ratio, $A_s$ is the amplitude of scalar fluctuations and $N_{ef}$ is the number of e-folds between the initial value $\sigma_i$ and the final $\sigma_f$. The parameters $\epsilon$ and $\eta$ are the so called slow-roll parameters. The condition for slow-roll inflation to happen, is that they are sufficiently small ($|\epsilon|, \, |\eta|<<1$) for a certain range of values of the scalar field. Here we will compute everything in terms of the field $\phi$, using the chain rule and the relation between $\sigma$ and $\phi$ given by \eqref{eq:relphisigma}.

If we assume the start (of the end) of inflation is at $\phi_i=1.55$, and determine the final of inflation when one of the slow roll parameters becomes $1$ (in our case it was $\epsilon$), it gives us the value for the field $\phi_f=025$. For that interval, one can say that the field slowly rolls on the potential, and one has the following results for the inflationary observables
\begin{align}
\quad n_s &\simeq 0.965 \,\,\,\,\,\,\,\,\,\,\,\,\,\,\,\,\,\,\,\,\,\,\,\,\,\,\,\,\, r \simeq 0.005 \nonumber \\ 
    \quad A_s & \simeq 2.06 \times 10^{-9} \,\,\,\,\,\,\, N_{ef}  \simeq 66 \, ,
\label{eq:computedobservables}
\end{align}
 where $n_s$, $r$ and $A_s$ where computed at $\phi_i$, and $N_{ef}$ between $\phi_i$ and $\phi_f$. These values are in agreement with current observations made by \cite{Planck:2018jri}. 

 At this stage, it is important to emphasise that these results where obtained on the basis of small amount of fine tuning. Only the numerical value of two mass couplings ($m_0^f, \, m_0^v$) had to be fixed in order to obtain this result. 
 The non-trivial field dependence of the obtained effective potential, is the result of the interplay between the gravitational degrees of freedom and the scalar degrees of freedom throughout the renormalization group flow between the NGFP and the infrared. To put it in other words, its shape is a consequence of the quantization of gravity plus the scalar degree of freedom, and its UV completion. 
 
 We have achieved what we claimed at the beginning of this work, and obtained, from a UV completion of gravity, an effective potential for a scalar degree of freedom that is consistent with a period of cosmological inflation. This potential is derived, not constructed, from fundamental physics, using only Wilsonian renormalization group techniques. 

 Up to now, every result was independent of any identification of the coarse graining scale with a physical scale, such as  the curvature of spacetime, or with the age of the universe, as is often done in the so called ``RG improvement'' procedures. All our results were obtained exactly in the limit of vanishing regulator. Nevertheless, one can take one more step. In the following, we will accept that one can relate the renormalization group scale $k$ with some physical scale, such that there is an identification between $k$ and physical time. This identification should be such that large values of $k$ are associated with early times of the universe, and small values of $k$ with later times. We do not need any particular identification, since our discussion does not require it.

\begin{figure}[h]
    \centering
    \includegraphics[width=\columnwidth]{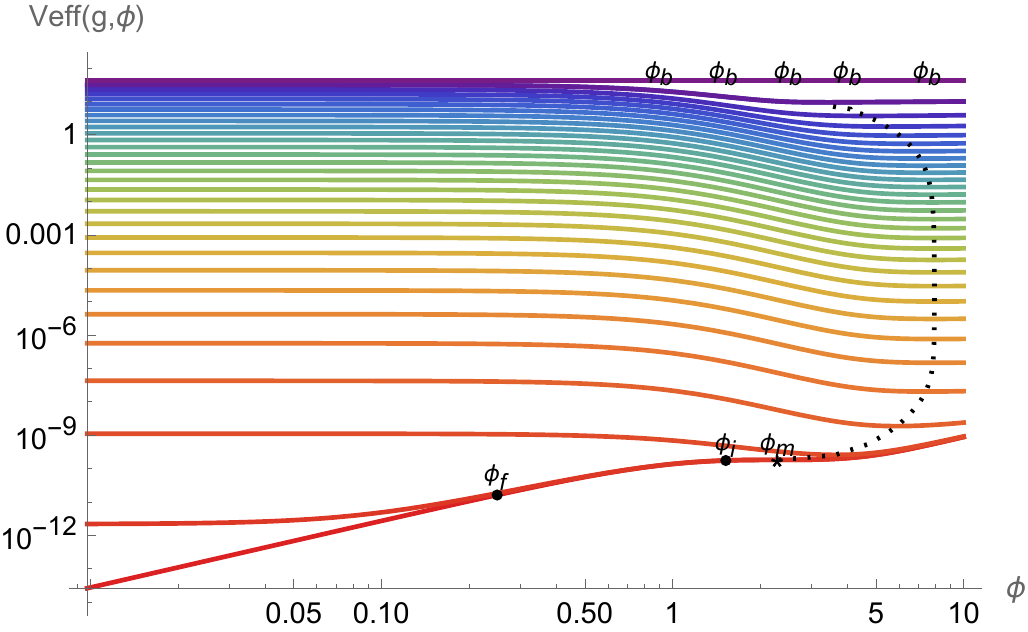}
    \caption{Snapshots of the RG evolution of $V_{eff}$, from the UV to the IR in logarithmic scale. The adiabatic evolution of the universe is shown by the dashed line, where it follows a stable minimum from the UV, that becomes unstable near $k \simeq 0$, around $\phi_m \simeq 2.3$.}
    \label{fig:ScalingVEFFCompleteCosmicHistory}
\end{figure}

 If one accepts that there is such identification, an argument that has been largely used \cite{Falls:2010he,Bonanno:2015fga,Bonanno:2012jy,Saueressig:2015xua,Koch:2013owa,Bonanno:2017pkg,Platania:2023srt}, but also very criticised \cite{Donoghue:2019clr,Bonanno:2020bil}, one can provide an explanation for the initial value of the inflaton field near the end of inflation. In Fig. \ref{fig:ScalingVEFFCompleteCosmicHistory}, one can see snapshots of the RG flow of the effective potential $V_{eff}$ as a function of $\phi$, in the studied range. Interestingly, the potential flows from the scale-invariant, and shift symmetric NGFP, where it takes the value $\lim_{k \to \infty} V_{eff}= 2 \lambda_{*}$, given by \eqref{eq:NGFPsplit1}, to the inflaton potential shown in Fig. \ref{fig:InflatonPotRegion}.

A tantalizing idea, extracted from the RG evolution of $V_{eff}$, is that if the very early universe was effectively described by an action close to \eqref{eq:scalartensoransatz} near the NGFP, the initial conditions for the inflaton field don't have to be fine tuned. This is because the NGFP is scale invariant, so the action at that fixed point will be independent of the value of the scalar field, and therefore the universe can have arbitrary initial conditions ($\phi_b$) for the field $\phi$ at its beginning. Furthermore, for values of $k<\infty$, $V_{eff}$ develops a stable minimum as a function of the scalar field $\phi$. If the universe begins near the NGFP, and is let to evolve adiabatically, the scalar field values will follow that minimum. This minimum becomes unstable near $k \simeq 0$, around $\phi_m \simeq 2.3$. This value is slightly above $\phi_i$, so that the field can roll down to $\phi_i$, where we start seeing the primordial fluctuations. Between $\phi_m$ and $\phi_i$ the universe experiences more than 70 e-folds, thus this region is not observable. An adiabatic evolution of the universe, with respect to the RG flow, will follow this trajectory and might explain the initial conditions of the inflaton near the end of inflation, without the need to fine tune its initial value, since it can start at an arbitrary value in the UV ($\phi_b$).

In Fig. \ref{fig:ScalingVEFFCompleteCosmicHistory}, snapshots of such evolution are presented, where the dashed line represent the values of the scalar field if the universe evolves adiabatically, with an arbitrary starting value of the scalar field $\phi_b$ at the beginning of the universe. Once all quantum fluctuations are integrated out, the stable minimum becomes unstable, and the end of inflation begins, giving rise to the observed fluctuations.

These and similar arguments have been used in several works before to describe the phenomenology of the RG flow, in particular when using RG improvements techniques, but also in other approaches, like using scale invariant solutions of the Wetterich equation \cite{Wetterich:2019qzx,Wetterich:2022ncl,Wetterich:2022brb,Wetterich:2019rsn}. The underlying justification for the assumption that the universe at its beginning is described by the highest energy degrees of freedom of the path integral ($\Gamma_k$ for large $k$) relies on the interpretation of $\Gamma_k$. Since at large values of $k$, $\Gamma_k$ contains information only of the highest energy modes of the theory, it is reasonable to believe that a denser universe in its beginnings will be mostly described by those degrees of freedom. As the universe evolves in time, it expands and becomes less dense, thus requiring the inclusion of lower energy modes for an accurate effective description ($\Gamma_k$ for lower values of $k$). We would like to emphasise that we did not use any particular identification of $k$ with a physical quantity, and if one desires to study details of the time evolution of the universe using Fig. \ref{fig:ScalingVEFFCompleteCosmicHistory}, one should stick to a particular identification, such as $k \propto \frac{1}{t}$ or something more sophisticated.

\section{Conclusions and outlook}\label{section:conclusions}
In this work, we studied the RG flow of a scalar tensor model given by \eqref{eq:scalartensoransatz}, with two arbitrary potentials for the scalar field. We showed the existence of an UV fixed point, with a finite number of relevant directions, and we used that fixed point to integrate the RG flow towards the limit of vanishing regulator, where one can make physical predictions. 

By means of this integration, we were able to predict the interaction potentials for the scalar field, fixing only two degrees of freedom, called $m_0^f$ and $m_0^v$ (see section \ref{section:numericalsol}). These two degrees of freedom span a continuum $2$-dimensional subspace of theory space where the Asymptotic Safety hypothesis is realized. 

By fixing $m_0^f=0.42$ and $m_0^v=2.75 \times 10^{-10}$, we determined numerically the effective potentials of the scalar field up to field values of $10$ times the Plank mass (see Fig. \ref{fig:predictedfandv}). Furthermore, by means of a conformal change of variables, going form the Jordan to the Einstein frame, we showed that our calculations give rise to an effective potential that is compatible with a slow roll inflationary period, driven by a single scalar field. The slow-roll observables obtained from such effective potential are compatible with current observations, giving the results $n_s \simeq 0.965, \, r\simeq 0.005, N_{ef}\simeq 66,\, A_s \simeq 2.06 \times 10^{-9}$ (see Fig. \ref{fig:InflatonPotRegion}). 

Furthermore, since the effective potential is shift-symmetric in the UV, we showed that a universe that evolves adiabatically towards the IR, and starts near the NGFP, does not require a fine tuning of the initial condition of the inflaton field. In fact, following such evolution, we determined that the inflaton field reaches an inflection point of the potential slightly above the value necessary to get more than 60 e-folds at the end of inflation, from where the scalar field can slowly roll down to the true minimum, giving rise to the observed fluctuations (see Fig. \ref{fig:ScalingVEFFCompleteCosmicHistory}). 

Our work leaves room for further improvement. All of our calculations of the RG flow were done in euclidean signature of the metric, and we could only compute physical quantities because we assumed a Bianchi type-I space-time, where a Wick rotation can be performed. Furthermore, the used RG flow equations, rely on a particular gauge choice, and a particular infrared regulator. On top of that, those equations are an approximation of the full flow equations, where one neglects the RG scale derivatives of the potentials in some sectors of the full flow equations. As mentioned before, a detailed discussion of such gauge elections, regulator and approximations can be found in \cite{Percacci:2015wwa}. It would be interesting to see if our results continue to hold varying such choices, for example using Lorentzian flows and the ADM formalism. If this is the case, these results would be compelling evidence in favor of Asymptotic Safety scenario for gravity.

\section{Acknowledgements}
The author is especially thankful to Frank Saueressig, for enlightening discussions, and for pinpointing key details of this work. The author also appreciates useful comments from Renate Loll, Cristóbal Laporte, and Jian Wang during the development of this work. 

\section{Appendix}{\label{sec:appendix}}
\subsection{Marginally relevant direction}{\label{sec:appendixreldirbs}} 
In order to determine if the direction \eqref{eq:reldirbreaksimmetry} is relevant at higher order in perturbations around the fixed point, we investigated it within a truncation of the action. In particular, if we assume that 
\begin{equation}
     F(k,\varphi) \simeq F0(k)+F1(k) \varphi^{2} \,\,\,\,\, V(k,\varphi) \simeq V0(k)+V1(k) \varphi^{2} \, ,
     \label{eq:truncatedexpansion}
\end{equation}
we can expand the flow equations \eqref{eq:flowFPercacci} and \eqref{eq:flowVPercacci} up to order $\varphi^2$, and obtain the flow equations for the couplings $F0(k)$, $V0(k)$, $F1(k)$ and $V1(k)$. In particular, the perturbation direction \eqref{eq:reldirbreaksimmetry} can be translated in this truncation into the perturbation $F0(k) \simeq F_* - \delta F0(k)$ and $F1(k) \simeq 32 \pi \delta F1(k)$, and fixing the other couplings to their fixed point value $V0_*=V_*,\, V1_*=0$, see \eqref{eq:NGFP}. Such perturbation satisfies the flow equations
\begin{align}
    k\,\frac{\partial\, \delta F0(k)}{\partial k} & =-2 (\delta F0(k)-\delta F1(k))\, , \nonumber \\
    k\,\frac{\partial\, \delta F1(k)}{\partial k} & =-\frac{3072 \pi ^2 \text{$\delta F1(k)$}^2 }{41}\, ,
\end{align}
with solutions 
\begin{align}
    \delta F1(k)& =\frac{41}{3072 \pi ^2 \log \left(\frac{k}{\text{k1}}\right)}\, , \nonumber \\
    \delta F0(k) &=\frac{\text{k0}^2}{k^2}-\frac{41 \text{k1}^2 \text{Ei}\left(2 \log
   \left(\frac{k}{\text{k1}}\right)\right)}{1536 \pi ^2 k^2}\, ,
\end{align}
where $k0$ and $k1$ are finite positive real numbers, $\text{Ei}$ is the exponential integral, and because we are expanding near the NGFP, we assumed $k>>k0,\,k1$. One can notice that $\lim_{k \to \infty} \{\delta F0(k),\,  \delta F1(k) \} = \{0, \,0 \}$, thus making the perturbation \eqref{eq:reldirbreaksimmetry} relevant. 

\subsection{Different boundary conditions}{\label{sec:appendixboundarycond}}

\begin{figure}
    \centering
    \includegraphics[width=\columnwidth]{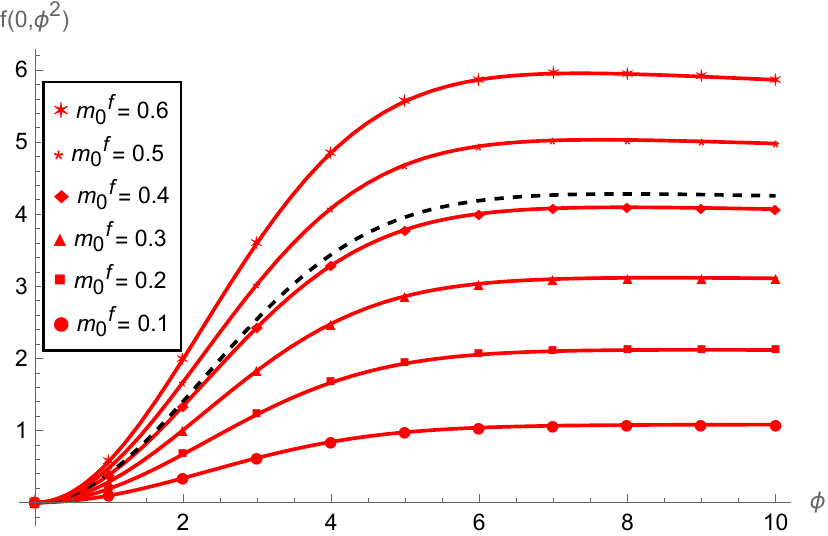}
    \caption{Effective potential $f$ in the limit of vanishing regulator ($k=0$) for different choices of the boundary condition parameter $m_0^f$, and fixing $m_0^v=2.75\times 10^{-10}$. With dashed lines is the reference potential used in this paper, corresponding to $m_0^f=0.42$.}
    \label{fig:diffboundarycondf}
\end{figure}

\begin{figure}
    \centering
    \includegraphics[width=\columnwidth]{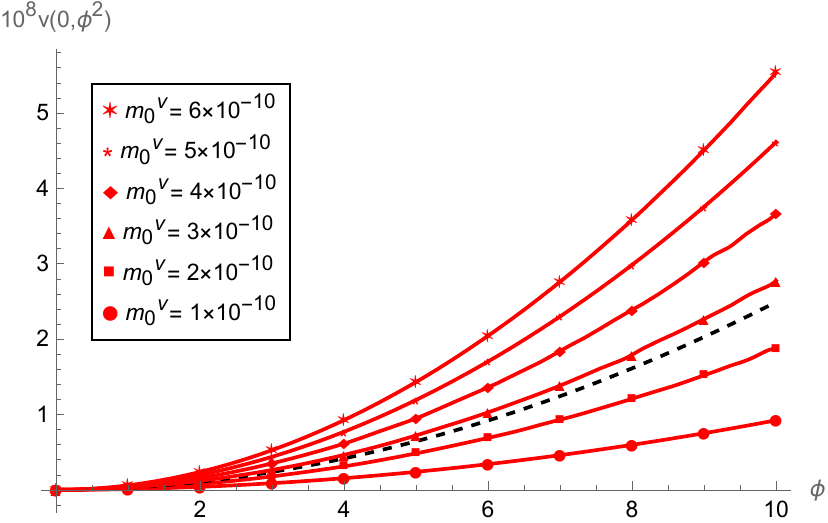}
    \caption{Effective potential $v$ in the limit of vanishing regulator ($k=0$) for different choices of the boundary condition parameter $m_0^v$, and fixing $m_0^f=0.42$. With dashed lines is the reference potential used in this paper, corresponding to $m_0^v=2.75\times 10^{-10}$.}
    \label{fig:diffboundarycondv}
\end{figure}

As was mentioned in the bulk of the paper, the results obtained in Figs. \ref{fig:predictedfandv}, and therefore in \ref{fig:InflatonPotRegion} and \ref{fig:ScalingVEFFCompleteCosmicHistory} are subject to a particular choice for the free parameters $m_0^f$ and $m_0^v$. Different choices for these parameters lead to different outcomes, and its effect in the obtained potentials is non-trivial. In Fig. \ref{fig:diffboundarycondf} and \ref{fig:diffboundarycondv} you can observe the result of changing these parameters to different values on the predicted potentials $f$ and $v$.

One can see that the effect of changing the values of the parameters $m_0^f$ and $m_0^v$ seems to be mostly an overall multiplication factor, since the shape of these potentials are very similar so each others. We do not have a proof of this conjecture, but it would be interesting to determine of this holds on a more formal setting than just looking at the results for some different values of $m_0^f$ and $m_0^v$. 

Not all the analysed values give rise to an effective potential in the Einstein frame that is compatible with observations, although, one can find other combinations apart from the ones showed in this paper that still work. The resulting effective potentials have different forms though, so their impact on other predictions, like primordial black holes, preheating and reheating can be different.

\subsection{Equations for the unknown scalar-field potentials}{\label{sec:appendixequations}}

Equations \eqref{eq:flowFPercacci} and \eqref{eq:flowVPercacci} are computed by keeping $\varphi$ independent of $k$, but we are interested in finding solutions in the variable $\phi$, keeping $\phi$ independent of $k$. This can be achieved by the following chain of replacements into equations \eqref{eq:flowFPercacci} and \eqref{eq:flowVPercacci}, applied in this specific order 
 \begin{align}
    k F^{(1,0)}(k,\varphi ) & \to \frac{ \beta_{g}}{g_k}\, \phi \, \mathcal{F}^{(0,1)}(k,\phi )+k \mathcal{F}^{(1,0)}(k,\phi ) \nonumber \\ 
      \quad F^{(0,n)}(k,\varphi ) &\to g_k^n \mathcal{F}^{(0,n)}(k,\phi ) \nonumber \\ 
      \quad k V^{(1,0)}(k,\varphi ) & \to \frac{ \beta_{g}}{g_k}\, \phi \, \mathcal{V}^{(0,1)}(k,\phi )+k \mathcal{V}^{(1,0)}(k,\phi ) \nonumber \\ 
      \quad V^{(0,n)}(k,\varphi ) &\to g_k^n \mathcal{V}^{(0,n)}(k,\phi ) \nonumber \\ 
      \quad \varphi  & \to \frac{\phi}{g_k},
      \label{eq:replacementsconstphi}
 \end{align}
  with $n=0,1,2$, $\beta_g= k \frac{\partial g_k}{\partial k}$, and $\mathcal{F}$ and $\mathcal{V}$ being intermediate auxiliary scalar functions. Once this is done, we will use the splitting given by \eqref{eq:splittingFandV}, that can be implemented by performing the following replacements
   \begin{equation}
     \mathcal{F}(k,\phi) \to \frac{1+f(g_k,\phi^{2})}{16 \pi g_k^2} \,\,\,\,\, \mathcal{V}(k,\phi) \to \frac{2\lambda(g_k) +v(g_k,\phi^{2})}{(16 \pi g_k^2)^{2}} \, ,
     \label{eq:replacementsconstphiplusgparamappendix}
 \end{equation}
and the corresponding partial derivatives.

On the resulting equations, one can determine $\beta_g$ and $\frac{\partial \lambda(g)}{\partial g}$ by evaluating each equation at $\phi=0$ and using that $f(g_k,0)=v(g_k,0)=0$. Once this is done, an interesting feature is that $\lambda$ decouples from the equations, leaving only partial differential equations for $f$ and $v$ as functions of $g$ and $\phi$.

As a result of these transformations, one ends up with two coupled, second order, non-linear, partial differential equations for $f(g,\phi^2)$ and $v(g,\phi^2)$, \eqref{eq:repflowv} and \eqref{eq:repflowf}. 

\begin{widetext}
\begin{align}
    &\frac{256 \pi  g^7 \left(4 \pi  (f(g,\phi^2 )+1) \left(v^{(0,1)}(g,\phi^2 )+2 \phi^2  v^{(0,2)}(g,\phi^2 )\right)-v^{(0,1)}(g,0) \left(3 \phi^2  f^{(0,1)}(g,\phi^2 )^2+4 \pi  (f(g,\phi^2 )+1)\right)\right)}{\left(v^{(0,1)}(g,0)+128 \pi ^2 g^2\right) \left(32 \pi  g^2 \left(3 \phi^2  f^{(0,1)}(g,\phi^2 )^2+4 \pi  (f(g,\phi^2 )+1)\right)+(f(g,\phi^2
   )+1) \left(v^{(0,1)}(g,\phi^2 )+2 \phi^2  v^{(0,2)}(g,\phi^2 )\right)\right)} \nonumber \\
   &-\frac{\left(24576 \pi ^3 g^7 f^{(0,1)}(g,0)-g \left(v^{(0,1)}(g,0)+128 \pi ^2 g^2\right) \left(\left(45 g^2-48 \pi \right) v^{(0,1)}(g,0)+128 \pi ^2 \left(41 g^2-48 \pi \right) g^2\right)\right) \left(2 v(g,\phi^2 )\right)}{24 \pi 
   \left(v^{(0,1)}(g,0)+128 \pi ^2 g^2\right)^2} \nonumber \\
   &+ \frac{\left(24576 \pi ^3 g^7 f^{(0,1)}(g,0)-g \left(v^{(0,1)}(g,0)+128 \pi ^2 g^2\right) \left(\left(45 g^2-48 \pi \right) v^{(0,1)}(g,0)+128 \pi ^2 \left(41 g^2-48 \pi \right) g^2\right)\right) \phi^2  v^{(0,1)}(g,\phi^2 )}{24 \pi 
   \left(v^{(0,1)}(g,0)+128 \pi ^2 g^2\right)^2} \nonumber \\
   & -\frac{g^{2} \left(\left(v^{(0,1)}(g,0)+128 \pi ^2 g^2\right) \left(\left(45 g^2-48 \pi \right) v^{(0,1)}(g,0)+128 \pi ^2 \left(41 g^2-48 \pi \right) g^2\right)-24576 \pi ^3 g^6 f^{(0,1)}(g,0)\right) v^{(1,0)}(g,\phi^2 )}{48 \pi  \left(v^{(0,1)}(g,0)+128 \pi ^2 g^2\right)^2} \nonumber \\
   &+g\left(-2 \phi^2 
   v^{(0,1)}(g,\phi^2 )+4 v(g,\phi^2 )\right)=0 
   \label{eq:repflowv}
\end{align}
\begin{align}
    &\frac{48 \pi  \phi^2  f^{(0,1)}(g,\phi^2 )}{g^2}+37-\frac{48 \pi  (f(g,\phi^2 )+1)}{g^2} \nonumber \\
    &-\frac{\phi^2  \left(24576 \pi ^3 g^6 f^{(0,1)}(g,0)-\left(v^{(0,1)}(g,0)+128 \pi ^2 g^2\right) \left(\left(45 g^2-48 \pi \right) v^{(0,1)}(g,0)+128 \pi ^2 \left(41 g^2-48 \pi \right) g^2\right)\right) f^{(0,1)}(g,\phi^2 )}{g^2 \left(v^{(0,1)}(g,0)+128 \pi ^2
   g^2\right)^2} \nonumber \\
   &-\frac{\left(\left(v^{(0,1)}(g,0)+128 \pi ^2 g^2\right) \left(\left(45 g^2-48 \pi \right) v^{(0,1)}(g,0)+128 \pi ^2 \left(41 g^2-48 \pi \right) g^2\right)-24576 \pi ^3 g^6 f^{(0,1)}(g,0)\right) f(g,\phi^2 )}{g^2 \left(v^{(0,1)}(g,0)+128 \pi ^2 g^2\right)^2} \nonumber \\
   &-\frac{\left(24576 \pi ^3 g^6
   f^{(0,1)}(g,0)-\left(v^{(0,1)}(g,0)+128 \pi ^2 g^2\right) \left(\left(45 g^2-48 \pi \right) v^{(0,1)}(g,0)+128 \pi ^2 \left(41 g^2-48 \pi \right) g^2\right)\right) f^{(1,0)}(g,\phi^2 )}{2 g \left(v^{(0,1)}(g,0)+128 \pi ^2 g^2\right)^2} \nonumber \\
   &-\frac{2 \left(v^{(0,1)}(g,0)+128 \pi ^2 g^2\right) \left(\left(45 g^2-48 \pi \right)
   v^{(0,1)}(g,0)+128 \pi ^2 \left(41 g^2-48 \pi \right) g^2\right)-49152 \pi ^3 g^6 f^{(0,1)}(g,0)}{2 g^2 \left(v^{(0,1)}(g,0)+128 \pi ^2 g^2\right)^2} \nonumber \\
   &+\frac{8 (f(g,\phi^2 )+1) \left(256 \pi ^2 g^4 \left(3 \phi^2  f^{(0,1)}(g,\phi^2 )^2+4 \pi  (f(g,\phi^2 )+1)\right) \left(-3 f^{(0,1)}(g,\phi^2 )-6 \phi^2  f^{(0,2)}(g,\phi^2 )+8 \pi \right)\right)}{\left(32 \pi  g^2 \left(3 \phi^2  f^{(0,1)}(g,\phi^2 )^2+4 \pi  (f(g,\phi^2 )+1)\right)+(f(g,\phi^2 )+1) \left(v^{(0,1)}(g,\phi^2
   )+2 \phi^2  v^{(0,2)}(g,\phi^2 )\right)\right)^2} \nonumber \\
   &+\frac{8 (f(g,\phi^2 )+1) \left(48 \pi
    g^2 \left(3 \phi^2  f^{(0,1)}(g,\phi^2 )^2+4 \pi  (f(g,\phi^2 )+1)\right) \left(v^{(0,1)}(g,\phi^2 )+2 \phi^2  v^{(0,2)}(g,\phi^2 )\right)\right)}{\left(32 \pi  g^2 \left(3 \phi^2  f^{(0,1)}(g,\phi^2 )^2+4 \pi  (f(g,\phi^2 )+1)\right)+(f(g,\phi^2 )+1) \left(v^{(0,1)}(g,\phi^2
   )+2 \phi^2  v^{(0,2)}(g,\phi^2 )\right)\right)^2} \nonumber \\
   &+\frac{8 (f(g,\phi^2 )+1) \left((f(g,\phi^2 )+1) \left(v^{(0,1)}(g,\phi^2 )+2 \phi^2  v^{(0,2)}(g,\phi^2 )\right)^2\right)}{\left(32 \pi  g^2 \left(3 \phi^2  f^{(0,1)}(g,\phi^2 )^2+4 \pi  (f(g,\phi^2 )+1)\right)+(f(g,\phi^2 )+1) \left(v^{(0,1)}(g,\phi^2
   )+2 \phi^2  v^{(0,2)}(g,\phi^2 )\right)\right)^2}=0 
   \label{eq:repflowf}
\end{align}
\end{widetext}
An implementation of these replacements can be found in the supplementary Mathematica notebook. Despite the complex nature of \eqref{eq:repflowv} and \eqref{eq:repflowf}, these equations are defined on a compact interval $g \in [0,\sqrt{\frac{48 \pi}{41}}]$, and can be solved numerically for any range of values of the scalar field $\phi$. They do not require any scaling with $k$, and from these equations one can directly extract the predictions in the limit of vanishing regulator ($k=0$) by evaluating the solutions at $g=0$.

\bibliography{bibliography}

\end{document}